\documentclass[twocolumn,aps,prb,superscriptaddress,a4paper,nofootinbib,showpacs,showkeys,lengthcheck]{revtex4-1}

\setcitestyle{numbers,square}

\usepackage{graphicx}
\usepackage{type1cm}
\usepackage[normalem]{ulem}

\usepackage{bm,dcolumn,amssymb,amsmath,braket,siunitx}

\newcommand{\mm}[1]{\mbox{$#1$}}
\newcommand{\dstd}{\mathrm{d}}

\newcommand{\ms}{\mbox{$\mu_{\mathrm{spin}}$}}
\newcommand{\mo}{\mbox{$\mu_{\mathrm{orb}}$}}

\newcommand{\Led}{\mbox{$L_{2,3}$\ edge}}
\newcommand{\ta}{\mbox{$T_{\alpha}$}}
\newcommand{\tz}{\mbox{$T_{z}$}}
\newcommand{\tth}{\mbox{$T_{\theta}$}}
\newcommand{\ef}{\mbox{$E_{F}$}}

\newcommand{\ea}{{\it et al.}}

\begin{document}

\title{Influence of spin-orbit coupling on the magnetic dipole term $T_{\alpha}$}



\author{O. \surname{\v{S}ipr}} 
\email{sipr@fzu.cz}
\homepage{http://www.fzu.cz/~sipr} \affiliation{Institute of Physics
  ASCR v.~v.~i., Cukrovarnick\'{a}~10, CZ-162~53~Prague, Czech
  Republic }

\author{J. \surname{Min\'{a}r}} \affiliation{Universit\"{a}t
  M\"{u}nchen, Department Chemie, Butenandtstr.~5-13,
  D-81377~M\"{u}nchen, Germany} \affiliation{New Technologies Research
  Centre, University of West Bohemia, Pilsen, Czech Republic}

\author{H. \surname{Ebert}} \affiliation{Universit\"{a}t M\"{u}nchen,
  Department Chemie, Butenandtstr.~5-13, D-81377~M\"{u}nchen, Germany}

\date{\today}

\begin{abstract}
The influence of the spin-orbit coupling (SOC) on the magnetic dipole
term \ta\ is studied across a range of systems in order to check
whether the \ta\ term can be eliminated from analysis of x-ray
magnetic circular dichroism spectra done via the spin moment sum rule.
Fully relativistic Korringa-Kohn-Rostoker (KKR) Green function
calculations for Co monolayers and adatoms on Cu, Pd, Ag, Pt, and Au
(111) surfaces were performed to verify whether the sum over magnetic
dipole terms $T_{x}+T_{y}+T_{z}$ is zero and whether the angular
dependence of the \ta\ term goes as $3\cos^{2}\theta-1$. It follows
that there are circumstances when the influence of the SOC on
\ta\ cannot be neglected even for 3$d$ atoms, where the SOC is
nominally small.  The crucial factor appears to be the dimensionality
of the system: for 3$d$ adatoms, the influence of SOC on \ta\ can be
significant while for monolayers it is always practically negligible.
Apart from the dimensionality, hybridization between adatom and
substrate states is also important: small hybridization enhances the
importance of the SOC and vice versa.
\end{abstract}



\maketitle


\section{Introduction}   \label{sec-intro}

Magnetism of diluted and low-dimensional systems such as adatoms,
clusters or monolayers is one of strongly pursued research areas.
Magnetization of these systems often cannot be measured by macroscopic
methods.  It can, however, be probed indirectly by making use of 
spectroscopy.  One of the most powerful methods in this respect is
x-ray magnetic circular dichroism (XMCD).  It consists in measuring
the difference in the absorption of left- and right-circularly
polarized x-rays in a magnetized sample while the energy of the
incident x-rays is varied.  Analysis of XMCD spectra is often done
with the help of sum rules, which link integrals of XMCD and x-ray
absorption spectral peaks to local spin and orbital magnetic moments.
Most of recent progress in magnetism of atomic-sized systems is
associated with the application of the XMCD sum rules
\cite{GRV+03,DSS+14,DRS+16}.

The strength of the sum rules is that they provide, in the case of
$L_{2,3}$ edge spectra,  separate
information about the orbital magnetic moment \mo\ and the spin
magnetic moment \ms\ of the photoabsorbing atom
\cite{TCSvdL92,CTAW93}.  However, extracting values of \mo\ and,
especially, of \ms\ from the spectra is not straightforward.  Considering
the most common case of the \Led\ spectra and a sample magnetized
along the $\alpha$ direction ($\alpha=x,y,z$), the spin magnetic
moment sum rule can be written as \cite{CTAW93}
\begin{equation}
 \frac{3}{I} \, \int \left( \Delta \mu_{{L}_3}
        -2\Delta \mu_{{L}_2} \right) \, \mathrm{d}E  \, = \, 
        \frac{\mu_{\text{spin}} + 7T_{\alpha}}{n_{h}}
   \enspace ,
\label{eq-spin}
\end{equation}
where \mm{\Delta \mu_{{L}_{2,3}}} are the differences
\mm{\Delta\mu=\mu^{(+)}-\mu^{(-)}} between absorption coefficients for
the left and right circularly polarized light propagating along the
$\alpha$ direction, $I$~is the integrated isotropic absorption
spectrum, $\mu_{\mathrm{spin}}$\ is the local spin magnetic moment
(its $d$\ component, to be precise), and $n_{h}$\ is the number of
holes in the $d$\ band. The term $T_{\alpha}$ is the expectation value
of the intra-atomic spin dipole operator for the valence $d$ electrons.  It
is often called the magnetic dipole term in the literature dealing
with XMCD.  As the magnetization is typically in the $\alpha=z$
direction, one often speaks simply about the \tz\ term.

This magnetic dipole \ta\ term can be written as \cite{Sto95,Sto99}
\begin{align}
T_{\alpha} & \: = \: - \frac{\mu_{B}}{\hbar} \, 
                     \langle \hat{T}_{\alpha} \rangle  
\;\;  \notag  \\
   & \: = \:  -\frac{\mu_{B}}{\hbar} \, 
   \left\langle \, \sum_{\beta} Q_{\alpha \beta} S_{\beta} \, 
      \right\rangle 
\enspace ,
 \raisetag{1.0\baselineskip}  \label{eq-exact}
\end{align}
with 
\begin{equation}
 Q_{\alpha \beta} \: = \: \delta_{\alpha \beta} \, - \, 
3 r^{0}_{\alpha} r^{0}_{\beta} 
\end{equation}
being the quadrupole moment operator and $S_{\alpha}$ being the spin
operator.  The \ta\ term cannot be easily determined by experiment and
its occurrence in Eq.~(\ref{eq-spin}) thus poses a serious
problem. For bulk systems, it can be often neglected (provided that
the spin-orbit coupling is not very strong \cite{CLT+95}).  However,
for low-dimensional systems it can be significant
\cite{WF94,KEDF02,SMC+10}.  Moreover, the \ta\ term cannot be
considered just as an additive correction that for similar systems simply
shifts the values of \ms\ by approximately the same amount.  It was
demonstrated that neglecting \ta\ for a sequence of supported magnetic
clusters could lead to erroneous conclusions regarding the dependence
of the average \ms\ on the cluster size \cite{SME09b}.  Likewise,
neglecting \ta\ and its angular dependence could introduce spurious
anisotropy of \ms\ for low-dimensional systems
\cite{Sto95,WSN+95,SBE+13}.

In principle, the \ta\ term can be calculated and inserted  into
Eq.~(\ref{eq-spin}).  However, one would really have to make the
calculation for each system which is studied,
because the \ta\ term is quite sensitive to details of the electronic
structure \cite{WF94,KEDF02,SME09b} and taking its values from
calculations for only similar systems might not be reliable. At the
same time, performing calculations for exactly the system one is
interested in may be difficult or impractical.

Fortunately, there appears to be a way to eliminate the \ta\ term from
Eq.~(\ref{eq-spin}) relying solely on experiment by performing a
series of angle-dependent XMCD measurements. The key here lies in
decoupling the quadrupole moment operator $\hat{Q}$ in
Eq.~(\ref{eq-exact}) from the spin operator $\hat{S}$. This can be
done provided that the influence of the spin-orbit coupling (SOC) on
\ta\ can be neglected.  Then, for a sample magnetically saturated
along the direction $\alpha$, one can express the \ta\ term as
\cite{SK95a}
\begin{equation}
T_{\alpha} \: = \:  \sum_{m} \, 
\frac{1}{2} \, 
\langle Y_{2m} |  \hat{Q}_{\alpha \alpha} | Y_{2m} \rangle \; 
\mu_{\text{spin}}^{(m)}
\;\; , 
\label{eq-tzspin}
\end{equation}
where $\mu_{\text{spin}}^{(m)}$ is the spin magnetic moment resolved
into components according to the magnetic quantum number $m$.  The
matrix elements $\langle Y_{2m} | \hat{Q}_{\alpha \alpha} | Y_{2m}
\rangle$ can be found in St\"{o}hr and K\"{o}nig \cite{SK95a} and more
elaborate discussion of Eq.~(\ref{eq-tzspin}) can be found in
St\"{o}hr \cite{Sto99} or \v{S}ipr \ea\ \cite{SBE+13}. Elimination of
the \ta\ term from the sum rule (\ref{eq-spin}) can then be achieved
by performing three XMCD measurements and making use of the relation
\cite{SK95a}
\begin{equation}
 T_{x} + T_{y} + T_{z} = 0 
\label{eq-sum}
\; .
\end{equation}
Furthermore, if the system has higher than twofold symmetry around the
$z$ axis, the magnetic dipole term depends on the polar angle $\theta$
as \cite{SK95a,Laan98}
\begin{equation}
 T_{\theta} \: \approx \: 3\cos^{2}\theta - 1
\label{eq-magic}
\; .
\end{equation}
The magnetic dipole term, which we will denote $T_{\theta}$ for a
general direction in which the sample is magnetically saturated, can
thus be eliminated by a single XMCD measurement with circularly
polarized x-rays coming in the direction of the magic angle
54.7$^{\circ}$.  This approach was employed, e.g., for studying Co
thin films and nanoclusters \cite{WSN+95,KMO+01}.

The important point is that eliminating \ta\ from the sum rule
analysis is possible only if the effect of SOC on \ta\ can be
neglected.  The question is whether this happens in common
circumstances. Namely, there are theoretical indications that the
effect of SOC on \ta\ may be sometimes important.  It was found that
Eq.~(\ref{eq-sum}) is strongly violated for free-standing Co wires
\cite{EKDF03} (provided that correlation effects beyond the LDA are
included via the Brooks orbital polarization term \cite{Bro85}).  For
more realistic materials, violation of Eq.~(\ref{eq-magic}) was
predicted for a Pt monolayer with magnetization induced from an Fe
substrate \cite{OS+04}.  Not surprisingly, this violation is even more
serious for systems with very strong SOC such as US
\cite{CLT+95,OS+04}.  Recently, there have been also experimental
indications that the SOC may be important for the \ta\ term: violation of
Eq.~(\ref{eq-sum}) was observed for low-temperature monoclinic phase
of magnetite nanoparticles \cite{SSW+14}.

The most typical situation when XMCD sum rules are used is studying
magnetism of 3$d$ metals in multicomponent systems, and the \ta\ term
has to be considered especially for thin films, adatoms or clusters.
One should thus explore to what extent the SOC is important for
\ta\ in these systems so that one knows whether
Eqs.~(\ref{eq-sum})--(\ref{eq-magic}) can be applied to eliminate the
\ta\ term from the XMCD analysis or not.

To get a comprehensive view, we focus on a sequence of systems
comprising Co monolayers and Co adatoms on Cu, Pd, Ag, Pt, and Au
(111) surfaces.  In that way we account for effects connected with the
change of dimensionality and for effects connected with the changes of
chemical environment as well.  It should be noted in this context 
that the substrate may have a crucial influence on some SOC-induced
properties such as the magnetocrystalline anisotropy \cite{SBE+14}.  There
is also theoretical evidence that the substrate has a decisive
influence on \ta\ of supported systems \cite{SME09b}.

The outline of the paper is the following.  We start by describing our
computational framework.  Then we present results that are in line
with 
Eqs.~(\ref{eq-sum})--(\ref{eq-magic}) for a series of Co monolayers
and adatoms.  Here we demonstrate that while for Co monolayers the
effect of SOC on \ta\ can be neglected for any of the investigated
substrates, the situation is complicated for Co adatoms, where for some
substrates Eqs.~(\ref{eq-sum})--(\ref{eq-magic}) are valid while for
others they are not.  This outcome is reinforced by inspection of the
validity of the approximate relation (\ref{eq-tzspin}) for \ta.
Finally, we investigate the density of states (DOS) to get an
understanding of why for adatoms on some substrates
Eqs.~(\ref{eq-sum})--(\ref{eq-magic}) are valid while for adatoms on
other substrates they are not.


\section{Computational scheme} 

 \label{sec-comput}

The calculations were performed within the ab-initio spin density
functional theory framework, relying on the local spin density
approximation (LSDA) with the Vosko, Wilk and Nusair parameterization
for the exchange and correlation potential \cite{VWN80}. The
electronic structure is described, including all relativistic effects,
by the Dirac equation, which is solved using the spin polarized
relativistic multiple-scattering or Korringa-Kohn-Rostoker (KKR) Green
function formalism \cite{EKM11} as implemented in the {\sc spr-tb-kkr}
code \cite{tbkkr-code}. The potentials were treated within the atomic
sphere approximation (ASA) and for the multipole expansion of the
Green function, an angular momentum cutoff
\mm{\ell_{\mathrm{max}}}=3 was used. The energy integrals were
evaluated by contour integration on a semicircular path within the
complex energy plane using a logarithmic mesh of 32 points.  The
integration over the $\mathbf{k}$ points was done on a regular mesh,
using 10000 points in the full surface Brillouin zone.

The electronic structure of Co monolayers on noble metals surfaces was
calculated by means of the tight-binding or screened KKR method
\cite{ZDU+95}. The substrate was modeled by a slab of 16 layers, the
vacuum was represented by 4 layers of empty sites.  The adatoms were
treated as embedded impurities: first the electronic structure of the
host system (clean surface) was calculated and then a Dyson equation
for an embedded impurity cluster was solved \cite{BMP+05}. The
impurity cluster contains 131 sites; this includes one Co atom, 70
substrate atoms and the rest are empty sites.  This cluster defines
the zone in which the electrons are allowed to react to the presence
of the adatom; there is an unperturbed host beyond this zone.

We investigate a series of Co adatoms and Co monolayers on the (111)
surface of the noble metals Cu, Ag, Au, Pd, Pt.  In this way we include
in our study substrates which are hard to magnetically polarize (Cu,
Ag, Au) and 
substrates that are easy to polarize (Pd, Pt), as well as substrates
with weak SOC (Cu), with moderate SOC (Pd, Ag), and with strong SOC
(Pt, Au).  We assume that all atoms are located on ideal lattice sites
of the underlying bulk fcc lattice; no structural optimization was
attempted.  While this would affect comparison of our data with
experiment, we do not expect this to have a significant influence on
the conclusions.


\section{Results}   \label{sec-res}


\subsection{Sum over magnetic dipole term components \ta}  

\label{sec-tzsum}

\begin{table}
\caption{Sum $7(T_{x} + T_{y} + T_{z})$ devided by \ms\ for Co
  monolayers and Co adatoms on noble metals surfaces.}
\label{tab-aver}
\begin{ruledtabular}
\begin{tabular}{ldd}
\multicolumn{1}{c}{substrate}   & 
    \multicolumn{1}{c}{monolayer} & 
    \multicolumn{1}{c}{adatom} \\
\hline 
Cu   &  0.011  &     0.206   \\
Pd   &  0.015  &     0.072   \\ 
Ag   &  0.021  &     0.372   \\
Pt   &  0.008  &     0.098   \\
Au   &  0.009  &     0.284  
\end{tabular}
\end{ruledtabular}
\end{table}

The first test of the influence of SOC on the \ta\ term is checking
the validity of Eq.~(\ref{eq-sum}). Our motivation comes from the spin
moment sum rule Eq.~(\ref{eq-spin}), in which \ms\ appears only in
combination with $7T_{\alpha}$, as $\mu_{\text{spin}}+7T_{\alpha}$.
The key indicator is thus the ratio $7T_{\alpha}/\mu_{\text{spin}}$.
Tab.~\ref{tab-aver} shows this ratio summed over all three
coordinates, $\sum_{\alpha=x,y,z}7T_{\alpha}/\mu_{\text{spin}}$.  It
should be zero if the influence of SOC on \ta\ can be neglected.

One can see that for Co monolayers the condition (\ref{eq-sum}) is
fulfilled with a high accuracy.  However, the situation changes for Co
adatoms.  It is obvious from Tab.~\ref{tab-aver} that the ratio
$\sum_{\alpha}7T_{\alpha}/\mu_{\text{spin}}$ is significantly larger
for adatoms than for the corresponding monolayers.  For Pd and Pt
substrates the breakdown of Eq.~(\ref{eq-sum}) is modest.  However,
for Cu, Ag, and Au substrates this breakdown is substantial.

The breakdown of Eq.~(\ref{eq-sum}) for adatoms is not related to any
specific choice of the coordinate system.  Similar numbers as those
shown in Tab.~\ref{tab-aver} are obtained if the sum over three
coordinate axes is substituted by a corresponding integral over the
full space angle (cf.\ also Fig.~\ref{fig-tz-theta} below).
It should be also noted that the dependence of the spin moment alone
on the magnetization direction is 
negligible: the variations do not exceed 0.03~\% for monolayers and
0.4~\% for adatoms.


\subsection{Angular dependence of magnetic dipole term}  

\label{sec-tzmagic}
 
\begin{figure}
\includegraphics[viewport=0.6cm 0.2cm 9.2cm 16.5cm]{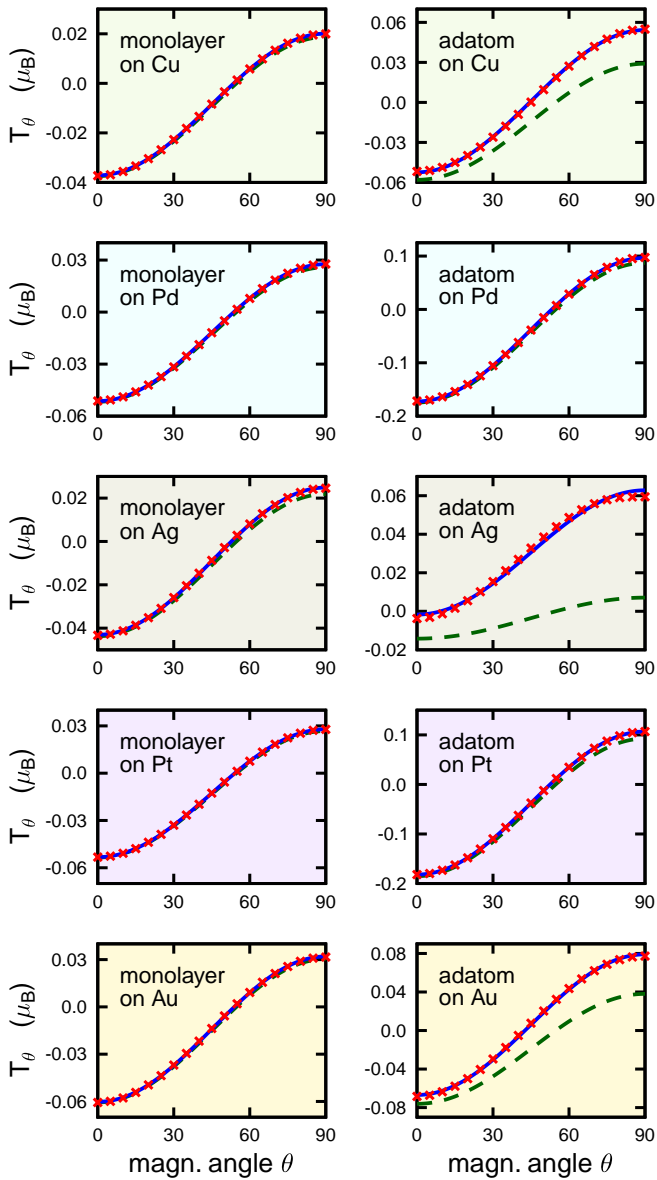}%
\caption{(Color online) Dependence of the magnetic dipole term
  \tth\ on the magnetization angle $\theta$ for Co monolayers (left
  panels) and Co adatoms (right panels) on different substrates.
  Ab-initio results are shown by red marks, fits to
  $A(3\cos^{2}\theta-1+B)$ are shown by full blue lines, fits to
  $A(3\cos^{2}\theta-1)$ are shown by dashed green lines.  Both fits
  are practically undistinguishable except for the cases of adatoms on
  Cu, Ag, or Au.}
\label{fig-tz-theta}
\end{figure}

Another view on the same problem can be obtained by inspecting the
angular dependence of the magnetic dipole term \tth.
Fig.~\ref{fig-tz-theta} shows the \tth\ term calculated while varying
the angle $\theta$ between the magnetization direction and the surface
normal.  The azimuthal angle $\phi$ was kept at \ang{0}, with the $x$
axis parallel to the [10$\bar{1}$] direction.  If the influence of SOC
can be neglected, the \tth\ dependence should satisfy
Eq.~(\ref{eq-magic}).  Therefore, we tried to fit our ab-initio data
to the expression 
\[ 
A \, (3\cos^{2}\theta-1) 
\]
(dashed green lines in Fig.~\ref{fig-tz-theta}).  This fit is
quite accurate except for Co adatoms on Cu, Ag, and Au.  In these
cases the \tth\ dependence can be fitted with the function
\[ 
A\, (3\cos^{2}\theta-1+B) 
\] 
(full blue lines in Fig.~\ref{fig-tz-theta}).

The fact that the \tth\ dependence can be fitted by
Eq.~(\ref{eq-magic}) only if a rigid shift (represented by the
constant $B$) is introduced presents another evidence that the
magnetic dipole term sum rule (\ref{eq-sum}) is not universally valid
for supported 3$d$ systems.  Likewise, \tth\ does not vanish at the
magic angle 54.7$^{\circ}$ for systems where $B$ is important.
Rather, it vanishes for a magnetization tilt angle of \ang{45} for a
Co adatom on Cu, \ang{13} for an adatom on Ag, and \ang{42} for an
adatom on Au.


\subsection{Approximate relation for \ta\ in terms of
  $\mu_{\text{spin}}^{(m)}$} 

\label{sec-mspin}

\begin{table}
\caption{Magnetic dipole term for $\bm{M}\|x$ ($T_{x}$) and
  $\bm{M}\|z$ ($T_{z}$) evaluated using the exact expression
  (\ref{eq-exact}) and using the approximate relation
  (\ref{eq-tzspin}).}
\label{tab-tz-approx}
\begin{ruledtabular}
\begin{tabular}{lcdddd}
   &  & 
    \multicolumn{2}{c}{Co monolayer} & 
    \multicolumn{2}{c}{Co adatom} \\ 
  \multicolumn{2}{l}{substrate}  & 
    \multicolumn{1}{c}{exact} & 
    \multicolumn{1}{c}{approx.} & 
    \multicolumn{1}{c}{exact} & 
    \multicolumn{1}{c}{approx.} \\ 
\hline 
Cu & $T_{x}$  &  0.020  &  0.021   &  0.057  &  0.031  \\
   & $T_{z}$  & -0.037  & -0.042   & -0.052  & -0.061  \\ [0.7ex]
Pd & $T_{x}$  &  0.028  &  0.027   &  0.099  &  0.093  \\
   & $T_{z}$  & -0.051  & -0.055   & -0.173  & -0.187  \\ [0.7ex]
Ag & $T_{x}$  &  0.025  &  0.024   &  0.059  &  0.008  \\
   & $T_{z}$  & -0.043  & -0.048   & -0.004  & -0.016  \\ [0.7ex]
Pt & $T_{x}$  &  0.028  &  0.028   &  0.109  &  0.098  \\
   & $T_{z}$  & -0.053  & -0.055   & -0.184  & -0.196  \\ [0.7ex]
Au & $T_{x}$  &  0.032  &  0.032   &  0.080  &  0.040  \\
   & $T_{z}$  & -0.061  & -0.064   & -0.066  & -0.079  
\end{tabular}
\end{ruledtabular}
\end{table}

Getting an intuitive insight into the \ta\ term by relying on the
exact Eq.~(\ref{eq-exact}) is not easy.  The approximate
Eq.~(\ref{eq-tzspin}) is far better suited for this purpose.  It
presents \ta\ as a linear combination of orbitally-projected
components of the spin magnetic moment $\mu_{\text{spin}}^{(m)}$,
illustrating thus the frequently used interpretation of the magnetic
dipole term  as manifestation of the anisotropy of spin density
distribution.  Indeed, if all $m$-components of \ms\ are identical,
\ta\ is zero.

However, this view is transparent only if the effect of SOC on
\ta\ can be neglected.  Therefore we present in
Tab.~\ref{tab-tz-approx} a comparison between values of \ta\ obtained
by evaluating the exact Eq.~(\ref{eq-exact}) and by evaluating the
approximate Eq.~(\ref{eq-tzspin}).  We focus on two magnetization
directions, $\bm{M}\|x$ and $\bm{M}\|z$.  One can see that as concerns
Co monolayers, the approximate equation yields similar values as the
exact equation.  For Co adatoms, the agreement is worse and, again, it
depends on the substrate.  For adatoms on Pd and Pt, the validity of
the approximate equation is worse than for corresponding monolayers
but it is still acceptable.  However, for adatoms on Cu, Ag, and Au
the error of the approximate Eq.~(\ref{eq-tzspin}) reaches 50--100~\%.

Comparison of the exact and approximate values of $T_{x}$ and $T_{z}$
in Tab.~\ref{tab-tz-approx} can serve as another indicator of the role
of SOC for the magnetic dipole term.  The outcome of this analysis is
consistent with the conclusions based on inspection of
Eq.~(\ref{eq-sum}) in Sec.~\ref{sec-tzsum} and Eq.~(\ref{eq-magic}) in
Sec.~\ref{sec-tzmagic}. Namely, the influence of the SOC on the
\ta\ term can be neglected for monolayers on any substrate and for
adatoms on Pd and Pt, while it has to be taken into account when
dealing with \ta\ for adatoms on Cu, Ag, and Au.


\subsection{Density of states}    \label{sec-dos}


\subsubsection{Total spin-polarized DOS}    \label{sec-totdos}

\begin{figure*}
\includegraphics[viewport=0.2cm 0.2cm 17.5cm 13.5cm]{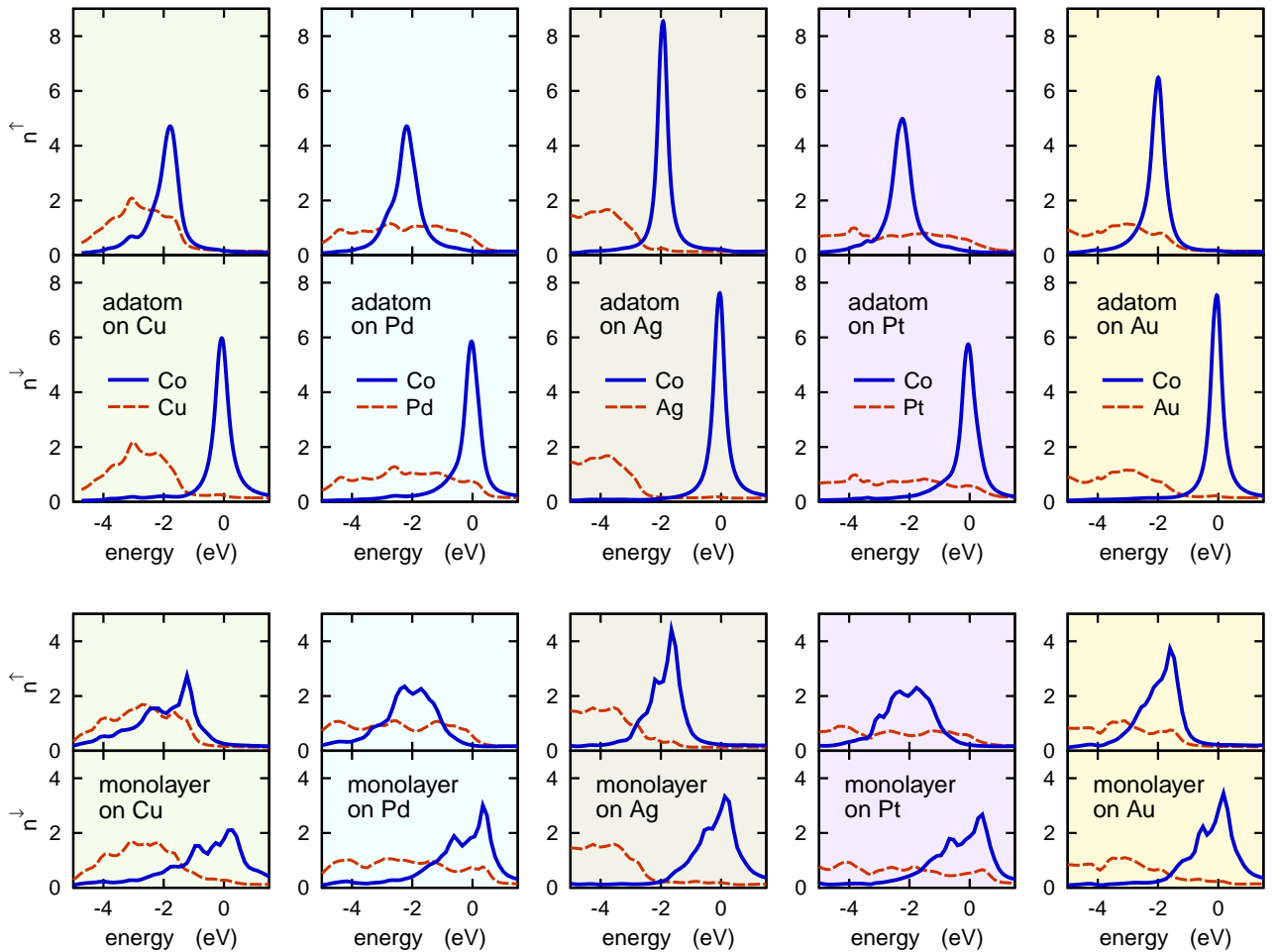}%
\caption{(Color online) Spin-polarized DOS for Co adatoms (upper
  panels) and monolayers (lower panels) on noble metals.  Blue solid
  lines show the DOS for Co atoms (in states per eV), dashed brown
  lines show the DOS for those substrate atoms which are nearest
  neighbors to Co atoms.}
\label{fig-DOS-overall}
\end{figure*}

To summarize, we found two trends concerning the impact of SOC on \ta.
First, the dimensionality or perhaps better the size of the system is
crucial: the effect of SOC can be always neglected for monolayers but
only sometimes for adatoms.  Second, there is a big variance depending
on the substrate but the nominal strength of the substrate SOC does
not seem to be important.

Thinking about the explanation, one should recall that the SOC
strength $\xi$ should be compared to the crystal field splitting
$\Delta_{\text{CF}}$ \cite{SK95a,SMP+16} --- it is the
\mbox{$\xi/\Delta_{\text{CF}}$} ratio that matters.  The splitting
$\Delta_{\text{CF}}$ is a model Hamiltonian parameter that is not
directly accessible by LDA calculations.  It can be seen as a measure
how electronic states around an atom are affected by the crystal field
due to its neighbors.  An idea how the influence of the crystal field
varies across our systems can be obtained by inspecting the DOS.
Therefore we present in Fig.~\ref{fig-DOS-overall} the spin-polarized
DOS for all the systems we investigate.  Apart from the DOS for Co
atoms we show also the DOS for the nearest substrate atoms, so that
hybridization between them can be studied.

One can see that (not surprisingly) the bandwidth for monolayers is
always significantly larger than the bandwidth for adatoms, no matter
what is the substrate. This clarifies why the influence of SOC on the
\ta\ term is negligible for the monolayers: in that case, the effect
of the crystal field always overwhelms the effect of SOC.


\subsubsection{DOS overlap integrals}

\begin{table}
\caption{Comparing the importance of SOC for \ta\ (characterized by
  sums over three $T_{\alpha}$ components, the second column) to the
  degree of atomic-like character of states associated with the adatom
  (characterized by reciprocal values of the DOS overlap integrals,
  the third colum).}
\label{tab-overlap}
\begin{ruledtabular}
\begin{tabular}{ldd}
   &  
    \multicolumn{1}{c}{relative weight of} & 
    \multicolumn{1}{c}{relative weight of} \\ 
   & 
    \multicolumn{1}{c}{$\sum_{\alpha}7T_{\alpha}/\mu_{\text{spin}}$} & 
    \multicolumn{1}{c}{$1 / \int \! \dstd E \, 
      n^{\downarrow}_{\text{Co}}(E) \, n^{\downarrow}_{\text{subs}}(E)$} \\ 
\hline 
Cu &  0.181  &  0.197   \\
Pd &  0.061  &  0.091   \\
Ag &  0.390  &  0.324   \\
Pt &  0.092  &  0.117   \\
Au &  0.276  &  0.269   \\
\end{tabular}
\end{ruledtabular}
\end{table}

What is not clear is why there are so big differences for the adatoms
when going from one substrate to another.  The bandwidth is
approximately the same for all substrates.  One can, nevertheless,
quantify the importance of hybridization between the adatom and the
substrate by evaluating the DOS overlap integral, i.e., the integral
of the product of the DOS for the adatom $n^{(s)}_{\text{Co}}$ and for
the nearest substrate atom $n^{(s)}_{\text{subs}}$,
\begin{equation}
 h^{(s)} \: \equiv \: \int \! \dstd E \, 
      n^{(s)}_{\text{Co}}(E) \, n^{(s)}_{\text{subs}}(E)
\; ,
\label{eq-dosint}
\end{equation}
where $s$ stands for the spin.  Here an interesting relation between
the integrals $h^{(s)}$ and the importance of the SOC for \ta\ appears
if we focus on the minority-spin states ($s=\downarrow$).  Namely, the
relative importance of SOC for \ta\ (quantified as the ratio
$\sum_{\alpha}7T_{\alpha}/\mu_{\text{spin}}$, cf.~Sec.~\ref{sec-tzsum}
and Tab.~\ref{tab-aver}) can be linked to the degree of atomic-like
character of minority-spin adatom states (quantified as
$1/h^{(\downarrow)}$). This emerges from Tab.~\ref{tab-overlap} where
relative weights of both quantities are shown.  It follows from
Tab.~\ref{tab-overlap} that if adatom states are less hybridized with
the substrate, indicating thus that the crystal field splitting is
smaller, the importance of SOC increases --- in agreement with
intuition.

The only caveat here is that this correspondence holds only for
minority-spin states; if majority-spin states are included in the
analysis, the correspondence between
$\sum_{\alpha}7T_{\alpha}/\mu_{\text{spin}}$ and $1/h^{(s)}$
disappears.  However, there is a reason for focusing on minority-spin
states only.  If majority-spin states are mostly occupied (as it is
the case for our systems), it is the incomplete occupancy of
minority-spin states which induces asphericity.  The importance of
partially-filled minority-spin states for \ta\ is also emphasized by
the fact that the value of \ta\ strongly depends on the position of
\ef, which cuts through minority-spin states
\cite{WF94,KEDF02,SME09b}.


\subsubsection{Orbitally-resolved DOS for Co adatoms}

\begin{figure*}
\includegraphics[viewport=0.2cm 0.2cm 17.5cm 9.5cm]{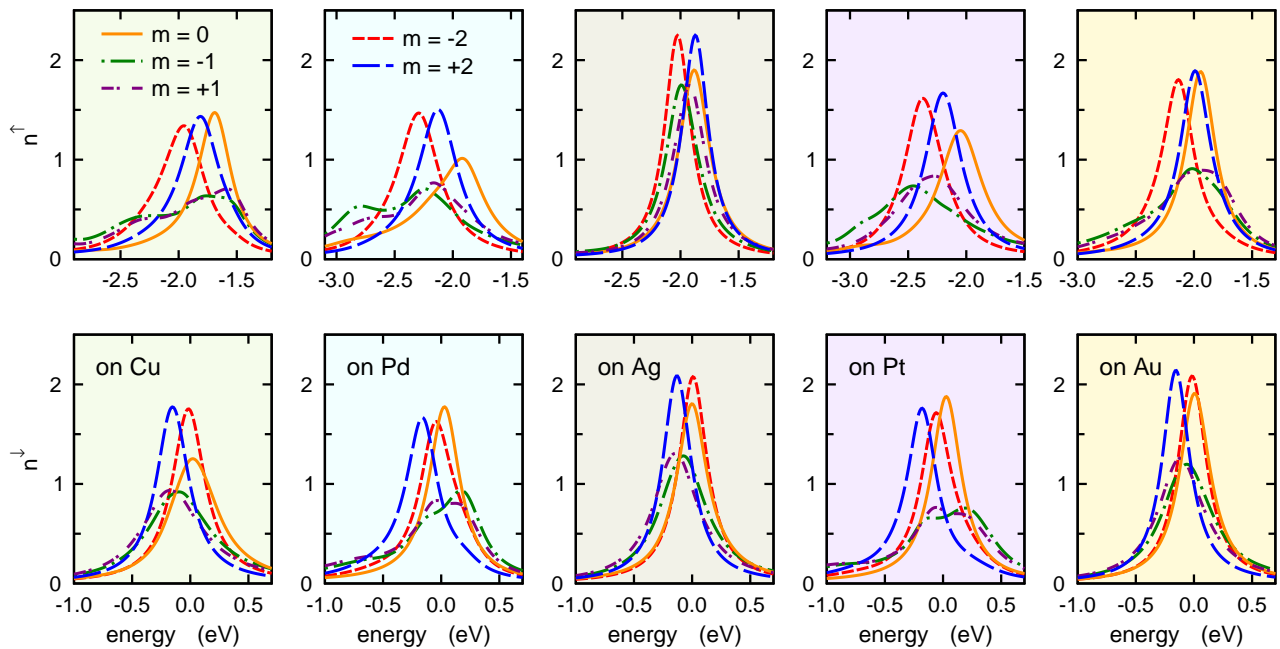}%
\caption{(Color online) Orbitally-resolved DOS (in states per eV) for
  Co adatoms on noble metals surfaces.  Majority-spin $d$ states are
  shown in the upper panels, minority-spin $d$ states are shown in the
  lower panels. Meaning of the lines is shown in the two upper left
  panels.}
\label{fig-mDOS}
\end{figure*}

Yet another view on hybridization of adatom states with substrate
states can be obtained from the orbitally-resolved DOS for the Co adatom.
This is shown in Fig.~\ref{fig-mDOS}: majority-spin states are
inspected in the upper panels, minority-spin states in the lower
panels. Several features in this plot are worth commenting.  First,
the 
individual orbital-resolved peaks are broader for the majority-spin states
than for this minority-spin states.  This is because for the majority-spin
states there is a considerable overlap with substrate states while for
the minority-spin states there is practically no overlap (see
Fig.~\ref{fig-DOS-overall}).  The Ag substrate with a deep lying $d$
band is an exception, in this
case the majority-spin adatom states have no overlap with substrate
states either.

The DOS peaks for $m=\pm2$ resemble broadened energy levels, as for an
isolated atom. This is because the orbital lobes for $m=\pm2$ lie
parallel to the surface, where there are no other atoms to hybridize
with.  The situation for $m=0$ is similar --- here the orbital lobe
points to the void between three nearest substrate atoms.  The
influence of the substrate is most pronounced for the $m=\pm1$
orbitals whose lobes are directed toward neighboring atoms.  Apart
from that, states for $m=\pm1$ and $m=\pm2$ are split by the SOC.
A more formal discussion about resolving the DOS according to spin and
orbital quantum numbers as well as about the role of the SOC-induced
splitting for the magnetocrystalline anisotropy was recently presented
by \v{S}ipr \ea~\cite{SMP+16}.

Here our attention is on the hybridization and, in particular, on the
difference between Cu, Ag, and Au substrates on the one hand and Pd
and Pt substrates on the other hand.  This difference is apparent for
the minority-spin states with $m=\pm1$ (lower panels of
Fig.~\ref{fig-mDOS}): while there is only one single peak for each of
the $m=\pm1$ components for the Cu, Ag, and Au substrates, there are
two peaks for the Pd and Pt substrates.  We can infer from this that
the crystal field splitting $\Delta_{\text{CF}}$ is small in the case
of a Co adatom on Cu, Ag, and Au, enabling thus the SOC to have a
large role for the \ta\ term, while it is large in the case of a Co
adatom on Pd and Pt, suppressing thus the role of SOC for \ta.  Anc
analysis of the orbital-resolved DOS thus reinforces the message
obtained by analyzing the overlap integrals (\ref{eq-dosint}) in
Tab.~\ref{tab-overlap}.


\section{Discussion}   \label{sec-discuss}

The purpose of this work was to study systematically the conditions
under which the influence of SOC on the \ta\ term can or cannot be
neglected for 3$d$ systems and, in this way, to explore possibilities
to eliminate the \ta\ term from the spin moment sum rule
(\ref{eq-spin}).  We found that even for atoms with low SOC such as Co,
the influence of SOC on \ta\ in certain environments can be so large
that Eqs.~(\ref{eq-sum})--(\ref{eq-magic}) cannot be used.  The
crucial factor turns out to be the ratio between SOC and crystal
field splitting, $\xi/\Delta_{\text{CF}}$.  This subsequently
translates itself into the dependence on the dimensionality.  It turns
out that for Co monolayers the influence of SOC on \ta\ can be
neglected for any of the Cu, Pd, Ag, Pt, or Au substrates.  We assume
that this is true for any 3$d$ monolayer on any substrate.

For adatoms the situation is more complicated.  The decrease of
$\Delta_{\text{CF}}$ caused by the decrease of the dimensionality
appears to be just of that size which is required for SOC to become
important for \ta.  Hence details of the electronic structure of the
substrate begin to matter; for some substrates (Pd, Pt)
Eqs.~(\ref{eq-sum})--(\ref{eq-magic}) still can be used while for
others (Cu, Ag, Au) they cannot.  The hybridization between adatom and
substrate states around \ef\ seems to be the deciding factor.  We
expect that for systems with considerable overlap between adatom and
substrate DOS around \ef\ (minority-spin states in our case, see
Fig.~\ref{fig-DOS-overall}) the influence of SOC on \ta\ can be
neglected even for adatoms.  Otherwise
Eqs.~(\ref{eq-sum})--(\ref{eq-magic}) should rather not be used.

To find more about when the size of the system gets so small that
Eqs.~(\ref{eq-sum})--(\ref{eq-magic}) cannot be used any more, we
performed calculations also for a Co wire on Au(111).  The wire was
built along the [1$\bar{1}$0] direction, we modelled it by a
2$\times1$ surface supercell.  To test whether Eq.~(\ref{eq-sum})
could be applied for such system, we evaluated the ratio
$\sum_{\alpha}7T_{\alpha}/\mu_{\text{spin}}$ and found it to be
0.058 \footnote{Our finding that SOC is not very important for
  \ta\ for a Co wire on Au(111) is not in contradiction with results
  of Ederer \ea\ \cite{EKDF03} because they studied a {\em
    free-standing} Co wire and, additionally, employed the Brooks
  orbital polarization correction.}. This is to be compared with 0.284
for a monolayer and 0.009 for an adatom (see Tab.~\ref{tab-aver}).  We
conclude, therefore, that the borderline between systems which satisfy
Eqs.~(\ref{eq-sum})--(\ref{eq-magic}) and which do not is somewhere
between the wire and the adatom.  When analyzing XMCD spectra for
small 3$d$ clusters of just few atoms, one should not rely on
Eqs.~(\ref{eq-sum})--(\ref{eq-magic}).  When analyzing XMCD spectra of
clusters of hundreds of atoms (as was the case, e.g., in the study of
Koide \ea\ \cite{KMO+01}), reliance on
Eqs.~(\ref{eq-sum})--(\ref{eq-magic}) is justified.


\section{Conclusions}   \label{sec-zaver}

The influence of spin-orbit coupling on the magnetic dipole term
\ta\ can be neglected for 3$d$ transition metal systems as long as
they are sufficiently large.  If the system contains just a few 3$d$
atoms (as is the case of adatoms or small supported clusters), the
influence of SOC on \ta\ may be significant.  This further depends on
the hybridization between states of the 3$d$ atoms and of the
substrate, especially around the Fermi level: if the hybridization is
only weak, the role of the SOC is enhanced while if the hybridization
is strong, the role of the SOC is suppressed.  For systems where the
influence of SOC on \ta\ cannot be neglected, the \ta\ term cannot be
eliminated from the XMCD spin sum rule --- neither by relying on the
$T_{x}+T_{y}+T_{z}=0$ relation, nor by making use of the magic angle
$\theta=54.7^{\circ}$.


\begin{acknowledgments}
This work was supported by the Ministry of Education, Youth and Sport
(Czech Republic) within the projects LD15097 (O.\v{S}.) and LO1402
(J.M.) and by the Deutsche Forschungsgemeinschaft within the project
SFB 689 ``Spinph\"{a}nomene in reduzierten Dimensionen''.
\end{acknowledgments}



\bibliography{liter-Tz-and-SOC}

\end{document}